\documentclass[journal]{IEEEtran}

\usepackage[latin1]{inputenc}
\usepackage{graphicx}
\usepackage{amssymb,amsmath,amsfonts,amsthm}
\usepackage{algorithm,algorithmic}
\usepackage{cite}
\usepackage{float}
\usepackage{graphics}
\usepackage{subfigure}
\usepackage{xspace}
\usepackage[usenames,dvipsnames]{color}
\usepackage{epsfig, hyperref}

\begin{document}

\title{SWIPT in 3-D Bipolar Ad Hoc Networks \\ with Sectorized Antennas}

\author{Ioannis Krikidis,~\IEEEmembership{Senior Member,~IEEE}
\thanks{I. Krikidis is with the Department of Electrical and Computer Engineering, University of Cyprus, Nicosia 1678 (E-mail: {\sf krikidis@ucy.ac.cy}).}
\thanks{This work was supported by H2020-MSCA-RISE under Grant 690750.}}

\maketitle

\begin{abstract}
In this letter, we study the simultaneous wireless information and power transfer (SWIPT) concept in 3-D bipolar ad hoc networks with spatial randomness. Due to three spatial dimensions of the network, we introduce a 3-D antenna sectorization that exploits the horizontal and the vertical spatial separation. The impact of 3-D antenna sectorization on SWIPT performance is evaluated for the power-splitting technique by using stochastic geometry tools. Theoretical and numerical results show that 3-D sectorization achieves a more accurate beam steering in the 3-D space, which can be beneficial for both the information and the power transfer.  
\end{abstract}

\vspace{-0.2cm}

\vspace{-0.2cm}
\section{Introduction}

\IEEEPARstart{S}{imultaneous} wireless information and power transfer (SWIPT) is a new communication paradigm, where wireless devices extract information and energy from the received radio-frequency (RF) signals \cite{RUI}. Due to practical limitations,  SWIPT  cannot  be  performed  from  the same signal without losses and practical implementations divide the  received  signal  in  two  parts; one  part  is  used for  information  transfer  and  another  part  is  used  for  power transfer \cite{KRI}. In this work, we focus on the power splitting (PS) technique, where the signal is split into two separate streams of different power \cite{RUI,KRI}. The time-splitting technique, where SWIPT is performed in the time domain,  is a special case of PS with a binary power splitting parameter and therefore is excluded in our analysis.

Due to the importance of path-loss degradation on SWIPT systems,  recent works study SWIPT from a large-scale point of view by taking into account the spatial randomness of the network deployment \cite{DIN}. In these works, the location of the devices and/or base stations is mainly modeled as a homogeneous Poisson point process (HPPP) in the 2-D space. To further boost the information/power transfer, directivity gains through antenna sectorization is introduced as a promising technology.  Sectorized antennas not only efficiently handle multi-user interference through spatial separation \cite{HUN}, but also facilitate wireless power transfer in environments with poor propagation characteristics \cite{KHA}. However, most of the work on SWIPT with/without sectorized antennas is limited to 2-D networks and does not consider recent 3-D  deployments.

A first effort to analyze the performance of a 3-D network is presented in \cite{PAN},  where the authors study the coverage probability for a 3-D large-scale cellular network with omnidirectional antennas. On the other hand, recent standardization activities (e.g, 3GPPP) focus on full-dimension multiple-input multiple-output technology, where antenna elements are placed into a 2-D planar array in order to achieve spatial separation to both the elevation and the traditional azimuth domains \cite{KIM,SIM}. The employment of 3-D antenna sectorization on 3-D networks with geometric randomness is a new research area without any previous work.   

In this letter, we study antenna sectorization for 3-D bipolar ad hoc networks with SWIPT-PS capabilities. Conventional 2-D sectorized models \cite{HUN} are extended to capture spatial separation in the horizontal domain as well as the vertical domain. The impact of 3-D sectorized antennas on the achieved information and power transfer is evaluated by using stochastic geometry tools. Closed-form expressions as well as asymptotic simplifications are derived for the  success probability and the average harvested energy.  We reveal that 3-D antenna sectorization is a useful tool  for 3-D networks and the associated additional degree of freedom (i.e., vertical spatial separation) achieves a higher information/power transfer than conventional antenna configurations. The main contributions of this letter are the development of a new 3-D antenna sectorization model for 3-D networks and its study in the context of SWIPT.

\vspace{-0.1cm}
\section{System model}\label{sec2}

We consider a 3-D Poisson bipolar ad hoc network consisting of a random number of transmitter-receiver pairs \cite{HUN}. The transmitters form a HPPP $\Phi=\{x_k \!\in\! \mathbb{R}^3\}$ with $k \geq 1$ of density $\lambda$ in the 3-D space \cite[Sec. 2.4]{HAN}, where $x_k$ denotes the coordinates of the node $k$. Each transmitter has a dedicated receiver (not a part of $\Phi$) at an Euclidean distance $d_0$ in some random direction.  The time is considered to be slotted and in each time slot all the sources  are  active  without any coordination or  scheduling  process. To analyze the performance of the network, consider a typical receiver located at the origin \cite{HAN}. This 3-D set-up is inline with ultra-dense urban networks, where a vast number of devices such as smartphones, tablets, sensors, connected objects will be connected in the same frequency band. These devices can be randomly spread in both the horizontal/vertical spatial domains in a chaotic manner.

We assume that wireless links suffer from both small-scale block fading and large-scale path-loss effects. The fading is Rayleigh distributed\footnote{More sophisticated 3-D channel models that capture the spatial correlation between the horizontal and the vertical domain are beyond the scope of this letter \cite{SIM}.} so the power of the channel fading is an exponential random variable with unit variance \cite{PAN}. We denote by $h_x$ the power of the channel fading for the link between the interfering transmitter $x$ and the typical receiver; $h_0$ denotes the power of the channel for the typical link. The path-loss model assumes that the received power is proportional to $1/(1+d^{\alpha})$ where $d$ is the Euclidean distance between the transmitter and the receiver, $\alpha>3$ denotes the path-loss exponent\footnote{The condition $\alpha>3$ ensures the validity of the derived closed-form expressions and holds for practical urban/suburban wireless environments. The  considered  path-loss model ensures  that  the path-loss is  always larger than  one for  any distance \cite{HAN} and is appropriate for SWIPT systems.}. In addition, all wireless links exhibit additive white Gaussian noise (AWGN) with variance $\sigma^2$. 
\begin{figure}[t]
\centering
 \includegraphics[width=0.87\linewidth]{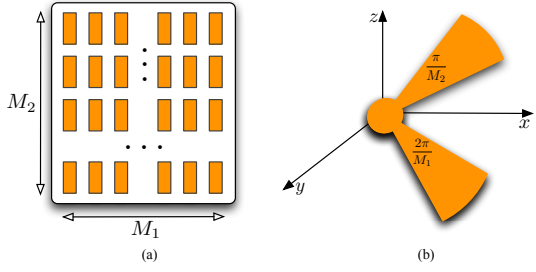}\\
\vspace{-0.9cm}
\caption{a) 2-D sectorized planar antenna array with $M_1\times M_2=M$ elements; b) Sector size for the horizontal domain ($2\pi/M_1$) and the vertical domain ($\pi/M_2$).}\label{model1}
\end{figure}

\vspace{-0.2cm}
\subsection{Azimuth and vertical sectorization (AVS)}

The transmitters are equipped with sectorized planar antenna arrays to steer their beams to specific directions in both the azimuth and the vertical domains \cite{KIM,SIM}. Each planar antenna array consists of $M_1\times M_2=M$  equally spaced antenna elements arranged in a regular rectangular array in the 2-D  plane; Fig. \ref{model1} schematically shows the planar antenna array and the associated azimuth/vertical sectorization. Each antenna element covers an angle $\frac{2\pi}{M_1}$ in the horizontal domain with directivity gain $M_1$, and an angle $\frac{\pi}{M_2}$ in the vertical domain with associated directivity gain $M_2$. We also assume a constant sidelobe level $\gamma$ for each antenna dimension with $0<\gamma< 1$ without loss of generality, where $\gamma$ denotes the ratio of the sidelobe level to the main lobe for the out-of-sector transmitted power for the horizontal and the vertical dimensions \cite{HUN}. The receivers are equipped with a single isotropic antenna and intercept RF electromagnetic fields equally in all possible directions in 3-D space. 

Due to the 3-D sectorization, three interference terms appear as follows: $\Phi_1$ represents the set of interferers which are aligned with the location of the typical receiver in the horizontal/vertical domains; $\Phi_1$ is a HPPP in the 3-D space with density $\lambda_1=\lambda\frac{1}{M_1}\frac{1}{M_2}=\frac{\lambda}{M}$ (thinning operation \cite{HAN}). $\Phi_2$ is the set of interferers which are aligned with the typical receiver only in the horizontal domain; $\Phi_2$ is a 3-D HPPP with density $\lambda_2=\lambda\frac{1}{M_1}\frac{M_2-1}{M_2}=\lambda\frac{M_2-1}{M}$. $\Phi_3$ refers to the set of interferers which transmit in different horizontal and vertical directions from the location of the typical receiver; it is also a 3-D HPPP with density $\lambda_3=\lambda\frac{M_1-1}{M_1}$. We note that the vertical domain is taken into account only when the interference link is aligned with the horizontal location of the typical receiver; if an interferer occurs in a misaligned horizontal sector then its vertical sector is also misaligned.

By using similar arguments with \cite{HUN}, Table \ref{table1} summarizes the interference characteristics for each of these processes, where $\lambda_i$ is the node density of the process $\Phi_i$, $\theta_i$ and $\phi_i$ denote the sector size over which the process occurs for the horizontal and the vertical domain, respectively, and $\psi_i$ is the total antenna directivity gain. As for the direct links, we assume that the transmitters are perfectly aligned with their associated receivers i.e., the antenna directivity gain equals to $\psi_1$.

\vspace{-0.2cm}
\subsection{SWIPT at the receivers}

The transmitters have their own power supply (e.g., battery) and transmit with full power $P_t$. The receivers employ a SWIPT-PS technique and thus split the received signal into two flows of different power for information decoding and energy harvesting, respectively. Let $0<\nu\leq 1$ denote the PS parameter  for  each  receiver \cite{KRI}. The energy harvesting circuit is ideal and the corresponding activation threshold is equal to zero \cite{RUI,DIN}. As for the information decoding process, an extra noise is introduced by the conversion operation of the RF band to baseband signal, which is modeled as AWGN with zero mean and variance $\sigma_C^2$. The  signal-to-interference-plus-noise ratio  (SINR)
at the typical receiver can be written as 
\begin{align}
{\sf SINR}=\frac{\frac{\nu P_t \psi_1 h_0}{\eta}}{\nu[\sigma^2+P_t(I_1+I_2+I_3)]+\sigma_C^2},
\end{align}
where $\eta\triangleq 1+d_0^\alpha$ and $I_i\triangleq\psi_i\sum_{x \in \Phi_i} \frac{h_x}{1+d_x^\alpha}$ denotes the (normalized) aggregate interference generated by the transmitters in $\Phi_i$.

\begin{table}[t]
\caption{Characteristics of the interference processes}
\centering
\resizebox{\columnwidth}{!}{
\begin{tabular}{|c||c||c||c||c|}
\hline
  - & $\lambda_i$ & $\theta_i$ & $\phi_i$ & $\psi_i$ \\
  \hline
 $I_1$ & $\lambda \frac{1}{M}$ & $2\pi\frac{1}{M_1}$ & $\pi \frac{1}{M_2}$ & $\frac{M}{(1+\gamma(M_1-1))(1+\gamma(M_2-1))}$ \\
 \hline
 $I_2$ & $\lambda \frac{M_2-1}{M}$ & $2\pi\frac{1}{M_1}$ & $\pi \frac{M_2-1}{M_2}$ & $\frac{\gamma M}{(1+\gamma(M_1-1))(1+\gamma(M_2-1))}$ \\
 \hline
 $I_3$ & $\lambda \frac{(M_1-1)}{M_1}$ & $2\pi\frac{M_1-1}{M_1}$ & $\pi$ & $\frac{\gamma^2 M}{(1+\gamma(M_1-1))(1+\gamma(M_2-1))}$\\
 \hline
\end{tabular}}\label{table1}
\end{table}

\vspace{-0.1cm}
\section{Performance analysis}

In this section, we study the information and power transfer performance of the 3-D bipolar ad hoc  network. 

\vspace{-0.3cm}
\subsection{Information transfer- success probability}

The information transfer performance of the network is analyzed in terms 
of  success probability i.e.,  the  probability that  the  receiver
can  support a  target  SINR threshold $\beta$. The probability of successful transmission for the typical receiver is 
\vspace{-0.1cm}
\begin{align}
\Pi_{\textrm{AVS}}&=\mathbb{P}\{{\sf SINR}\geq \beta \} \nonumber \\
&\;=\mathbb{P}\left\{h_0\geq \frac{\beta \eta [\nu(\sigma^2+P_t\sum_{i=1}^3 I_i)+\sigma_C^2]}{\nu \psi_1 P_t}  \right\} \nonumber \\
&\;=\exp\left(-\frac{\beta \eta (\nu\sigma^2+\sigma_C^2)}{\nu \psi_1 P_t}\right)\mathbb{E}_{\Phi} \exp\left(-\frac{\beta \eta \sum_{i=1}^3 I_i}{\psi_1} \right) \nonumber \\
&\;=\exp\left(-\frac{\beta \eta (\nu\sigma^2+\sigma_C^2)}{\nu \psi_1 P_t}\right)\prod_{i=1}^3\mathbb{E}_{\Phi_i}\! \exp\left(-\beta \eta I_i/\psi_1  \right) \nonumber \\
&\;=\exp\!\left(-\frac{\beta \eta (\nu\sigma^2+\sigma_C^2)}{\nu \psi_1 P_t}\right) \prod_{i=1}^3\mathcal{L}_I \left(\frac{\psi_i\beta \eta}{\psi_1},\lambda_i \right),  \label{res1} \\
\Pi_{\textrm{AVS}}^\infty &\;\rightarrow \mathcal{L}_I \left(\beta \gamma^2 \eta,\lambda \right)\;\;\;\text{for $P_t,M\rightarrow \infty$}, \label{ass1}
\end{align}
where $L_I(\cdot)$ denotes the Laplace transform of the interference and is given in the Appendix. 

The performance of conventional antenna configurations  yields from \eqref{res1} with appropriate parameterization.  Specifically, the case where the transmitters are equipped with single omnidirectional antennas (case ``OM'') \cite{PAN} refers to $M_1=M_2=1$ and is equal to
\begin{align}
\Pi_{\text{OM}}&=\exp\left(-\frac{\beta \eta(\nu\sigma^2+\sigma_C^2)}{\nu P_t}\right)\mathcal{L}_I(\beta\eta,\lambda), \\
\Pi_{\text{OM}}^\infty &\rightarrow \mathcal{L}_I(\beta\eta,\lambda)\;\;\;\text{for $P_t\rightarrow \infty$}.  \label{ass2}
\end{align} 
The case where the transmitters are equipped with a 1-D linear array of $M$ sectorized antennas for spatial separation on the traditional azimuth domain \cite{HUN} (case ``AS'') refers to $M_1=M$, $M_2=1$ and corresponds to   
\begin{align}
\Pi_{\text{AS}}&=\exp\left(-\frac{\beta \eta (\nu\sigma^2+\sigma_C^2)(1+\gamma(M-1))}{\nu M P_t}\right) \nonumber \\
&\;\;\;\;\times \mathcal{L}_I(\beta \eta,\lambda/M)\mathcal{L}_I(\beta\gamma \eta,\lambda(M-1)/M), \\
\Pi_{\text{AS}}^\infty &\rightarrow \mathcal{L}_I(\beta\gamma \eta,\lambda)\;\;\;\text{for $P_t,M\rightarrow \infty$}.  \label{ass3}
\end{align}
By comparing the asymptotic expressions in \eqref{ass1}, \eqref{ass2}, and \eqref{ass3}, we have $\Pi_{\text{OM}}^\infty<\Pi_{\text{AS}}^\infty<\Pi_{\text{AVS}}^\infty$. 

\vspace{-0.2cm}
\subsection{Power transfer- average harvested energy}

On  the  other hand,  RF  energy  harvesting  is  a  long  term  operation and is expressed in terms of average harvested energy \cite{RUI}.  If $100(1-\nu)\%$ of the received energy is used for rectification, the average energy harvesting at the typical receiver is expressed as
\begin{align}
{\sf E}_{\text{AVS}}&=\zeta \mathbb{E}_{\Phi,h} (1-\nu)\left[\frac{P_t \psi_1 h_0}{\eta}+P_t(I_1+I_2+I_3)  \right] \nonumber \\
&=\zeta (1-\nu)P_t\left(\frac{\psi_1}{\eta}+\sum_{i=1}^3 \psi_i\mathbb{E}_{\Phi_i} \sum_{x \in \Phi_i}\frac{1}{1+d_x^\alpha}  \right) \nonumber \\
&=\zeta (1-\nu)P_t\left(\frac{\psi_1}{\eta}+\sum_{i=1}^3\psi_i \mathcal{E}_I(\lambda_i) \right), \\
{\sf E}_{\text{AVS}}^\infty \!&\rightarrow\!{\zeta (1\!-\!\nu)P_t}\bigg(\!\frac{1}{\eta \gamma^2}\!+\!\frac{4\pi^2 \lambda}{\alpha\sin(\frac{3\pi}{\alpha})}  \!\bigg)\;\text{for $M_1,M_2\!\rightarrow\!\! \infty$}, \label{asym}
\end{align} 
\noindent where $\zeta \in (0 , 1]$ denotes  the  conversion  efficiency  from RF  signal  to  DC  voltage \cite{RUI,DIN}, and $\mathcal{E}_I(\cdot)$ is given in the Appendix. It  is  worth  noting  that  the  RF energy harvesting from the AWGN noise is considered to  be negligible. The expression in \eqref{asym} shows that the average harvested energy asymptotically depends on the ratio $\gamma$, while the ambient harvested energy (i.e., harvesting from the multi-user interference) is equal to the one of the OM and AS cases (see \eqref{exp1}, \eqref{exp11}).

For the special cases with $M_1=M_2=1$ (single omnidirectional antenna) and $M_1=M$, $M_2=1$ (sectorized linear antenna array), the average harvested energy is given by  
\begin{align}
{\sf E}_{\text{OM}}&=\zeta (1-\nu)P_t\left[\frac{1}{\eta}+\frac{4\pi^2\lambda}{\alpha \sin(\frac{3\pi}{\alpha})} \right], \label{exp1}\\
{\sf E}_{\text{AS}}&={\zeta (1-\nu)P_t}\bigg(\frac{M}{\eta[1+\gamma(M-1)]}+\frac{4\pi^2 \lambda}{\alpha\sin(\frac{3\pi}{\alpha})}  \bigg), \label{exp11} \\
{\sf E}_{\text{AS}}^\infty &\rightarrow {\zeta (1-\nu)P_t}\bigg(\frac{1}{\eta \gamma}\!+\!\frac{4\pi^2 \lambda}{\alpha\sin(\frac{3\pi}{\alpha})}  \bigg)\;\text{with $M\rightarrow \infty$}. \label{asym3}
\end{align}
By comparing the asymptotic expressions in \eqref{asym}, \eqref{exp1} and \eqref{asym3}, we have
${\sf E}_{\text{OM}}<{\sf E}_{\text{AS}}^\infty<{\sf E}_{\text{AVS}}^\infty$.

\section{Numerical results}

Computer simulations are carried-out in order to evaluate the performance of the proposed schemes. The simulation set-up follows the description in Section \ref{sec2} with parameters (unless otherwise defined) $P_t=-10$ dBm, $\alpha=4$, $\lambda=10^{-4}$, $d_0=3$ m, $M=16$, $M_1=M_2=\sqrt{M}$, $\beta=20$ dBm, $\sigma^2=-130$ dBm, $\sigma_C^2=-30$ dBm, $\nu=0.5$, $\zeta=1$, and  $\gamma=0.3$ \cite{LU}.  Conventional antenna configurations such as omnidirectional antennas (OM) and azimuth sectorization (AS) are used as benchmarks. In addition, existing 2-D models are presented for the OM and AS configurations \cite{PAN}.

\begin{figure}[t]
\centering
 \includegraphics[width=0.89\linewidth]{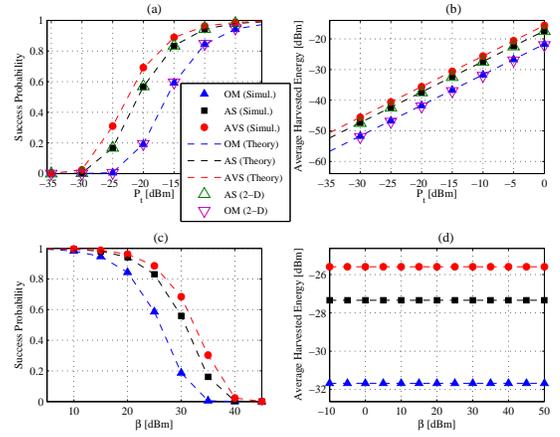}\\
\vspace{-0.3cm}
\caption{a) Success probability versus  $P_t$; b) Average harvested energy versus $P_t$; c) Success probability versus $\beta$; d) Average harvested energy versus $\beta$.}\label{fig1}
\end{figure}

Fig. \ref{fig1} plots the success probability and the average harvested energy versus the transmit power $P_t$ and the SINR threshold $\beta$. As it can be seen, antenna sectorization significantly facilitates the information and the power transfer in environments with low ambient interference. 3-D sectorization achieves a more accurate beam orientation by exploiting the horizontal as well as the vertical spatial separation and thus achieves higher success probability and more efficient energy transfer by boosting the direct links. The comparison with the conventional 2-D models shows the compatibility of the 3-D models and confirms the superiority of the 3-D antenna sectorization. Theoretical curves perfectly match with the numerical results and validate our analysis.

\begin{figure}[t]
\centering
 \includegraphics[width=0.9\linewidth]{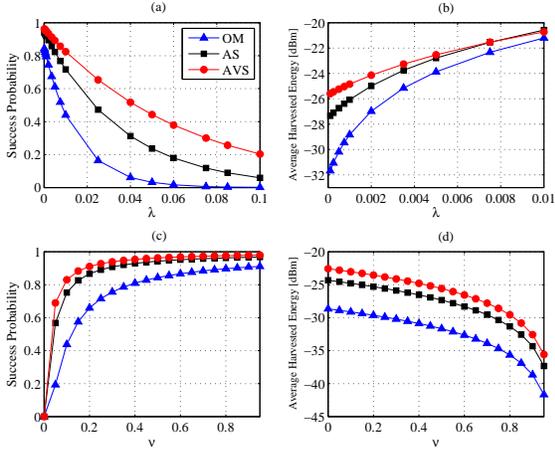}\\
\vspace{-0.4cm}
\caption{a) Success probability versus $\lambda$; b) Average harvested energy versus $\lambda$; c) Success probability versus $\nu$; d) Average harvested energy versus $\nu$.}\label{fig2}
\end{figure}

\begin{figure}[t]
\centering
 \includegraphics[width=0.9\linewidth]{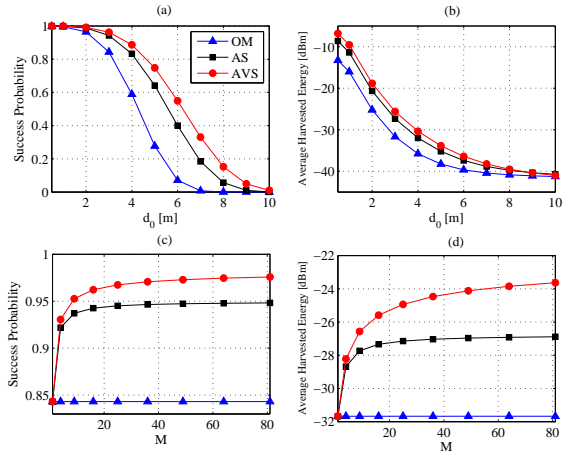}\\
\vspace{-0.4cm}
\caption{a) Success probability versus $d_0$; b) Average harvested energy versus $d_0$; c) Success probability versus $M$; d) Average harvested energy versus $M$.}\label{fig3}
\end{figure}

The impact of the network density and the PS parameter on the achieved information/power transfer is depicted in Fig. \ref{fig2}. Fig. \ref{fig2}.(a) shows that  sectorization is more beneficial for moderate network densities, where multi-user interference can be efficiently controlled through spatial separation. From the harvesting standpoint, Fig. \ref{fig2}.(b) illustrates that as $\lambda$ increases, multi-user interference becomes the dominant component of the harvested energy and thus antenna sectorization has limited interest.  On the other hand, Fig.'s \ref{fig2}.(c), \ref{fig2}.(d) capture the fundamental trade-off between information and energy transfer; as $\nu$ increases  the success probability  is  improved  while  the  harvesting  process  becomes less  efficient,  since  most  of  the  received  energy  is  used  for information decoding.
        
Finally, Fig. \ref{fig3} illustrates the impact of the Euclidean distance $d_0$ as well as the number of antenna elements $M$ on the system performance. As it can be seen from Fig.'s \ref{fig3} (a), \ref{fig3} (b), as the distance $d_0$ increases, the success probability decreases since the direct links become more weak. In addition, harvesting gains from antenna sectorization become less significant as $d_0$ increases, since the ambient multi-user interference dominates the harvesting process.  Fig.'s \ref{fig3} (c), \ref{fig3} (d) indicate that as the number of antenna elements increases, antenna steering becomes more accurate which is beneficial for information/power transfer through direct links; the convergence floor of the curves also validates the proposed asymptotic expressions.     

\vspace{-0.2cm}
\section{Conclusion}

In this letter, a new tractable 3-D model for bipolar ad hoc
networks with antenna sectorization has been presented and studied from a SWIPT point of view. Simple closed-form expressions for the success probability and the average harvested energy have been derived for the SWIPT-PS technique. Our results have shown that the 3-D antenna sectorization fully exploits the three spatial dimensions of the network and outperforms conventional antenna configurations, mainly at moderate network densities and transmitter-receiver distances.   
\vspace{-0.4cm}

\appendix
We derive the Laplace transform of the interference term $I=\sum_{x \in \Phi'}\frac{h_x}{1+d_x^\alpha}$, where $\Phi'$ is a thinned version of the process $\Phi$ with density $\xi\leq \lambda$. We have
\vspace{-0.2cm}
\begin{subequations}
\label{lap}
\begin{align}
\mathcal{L}_I(s,\xi)&=\mathbb{E}_{\Phi',h} \exp(-s I)\;=\;\mathbb{E}_{\Phi',h}\prod_{x\in \Phi'}\exp \left(-\frac{sh_x}{1+d_x^\alpha} \right) \nonumber  \\
&=\mathbb{E}_{\Phi'}\prod_{x \in \Phi'}\mathbb{E}_h \exp \left(-\frac{sh}{1+d_x^\alpha} \right) \nonumber \\
&= \exp\left(-4\pi \xi \int_{0}^{\infty} \left[1-\mathbb{E}_h \exp\left(-\frac{sh}{1+r^\alpha} \right) \right]r^2 dr \right) \label{lap0}   \\
&=\exp \left(-4\pi\xi \int_{0}^{\infty}\frac{sr^2}{1+s+r^\alpha} dr \right) \label{lap1}  \\
&=\exp \left(-\frac{4 \pi^2\xi}{\alpha \sin\left( \frac{3\pi}{\alpha} \right)} s(1+s)^{\frac{3-\alpha}{\alpha}} \right) \label{lap2},
\end{align}
\end{subequations}
where \eqref{lap0} uses the probability generating functional of a PPP and the mapping of a 3-D PPP to one dimension (with intensity function $\xi'(r)=4\pi\xi r^2$) \cite{HAN}, \eqref{lap1} follows from the moment generating function of an exponential random variable  i.e., $h_x\sim \exp(1)$, and  \eqref{lap2} is based on \cite[Eq. 3.241.4]{GRAD}.

In addition, we derive the mean interference over the HPPP $\Phi'$ as follows  
\begin{subequations}
\label{mean}
\begin{align}
\mathcal{E}_I(\xi)&=\mathbb{E}_{\Phi',h}\sum_{x \in \Phi'} \frac{h_x}{1+d_x^\alpha}=4\pi \xi \int_{0}^{\infty}\frac{r^2}{1+r^\alpha} dr \label{mean1} \\
&=\frac{4\pi^2 \xi}{\alpha\sin(\frac{3\pi}{\alpha})}, \label{mean2}
\end{align}
\end{subequations}
where \eqref{mean1} uses the Campbell's theorem for sums \cite{HAN}, and \eqref{mean2} is based on \cite[Eq. 3.241.4]{GRAD}.

\end{document}